# Superconductivity in Li$_6$P electride

Ziyuan Zhao[1], Shoutao Zhang[1], Tong Yu[1], Haiyang Xu[1], Aitor Bergara[*,2,3,4] and Guochun Yang[*,1]

Electrides are unique compounds where most of the electrons reside at interstitial regions of the crystal behaving as anions, which strongly determines its physical properties. Interestingly, the magnitude and distribution of interstitial electrons can be effectively modified either by modulating its chemical composition or external conditions (e.g. pressure). Most of the electrides under high pressure are non-metallic, and superconducting electrides are very rare. In this work we report that a pressure-induced stable Li$_6$P electride becomes superconductor with a $T_c$ of 39.3 K, which is the highest among already known electrides. The interstitial electrons in Li$_6$P, with dumbbell-like connected electride states, play a dominant role in the superconducting transition. Other Li-rich phosphides, Li$_5$P and Li$_8$P, are also predicted to be superconducting electrides, but with a lower $T_c$. Superconductivity in all these compounds can be attributed to a combination of a weak electronegativity of P with a strong electropositivity of Li, and opens up the interest to explore high-temperature superconductivity in similar binary compounds.

[1]Centre for Advanced Optoelectronic Functional Materials Research and Laboratory for UV Light-Emitting Materials and Technology of Ministry of Education, Northeast Normal University, Changchun 130024, China. [2]Departamento de Física de la Materia Condensada, Universidad del País Vasco-Euskal Herriko Unibertsitatea, UPV/EHU, 48080 Bilbao, Spain. [3]Donostia International Physics Center (DIPC), 20018 Donostia, Spain. [4]Centro de Física de Materiales CFM, Centro Mixto CSIC-UPV/EHU, 20018 Donostia, Spain. Correspondence and requests for materials should be addressed to a.bergara@ehu.eus; yanggc468@nenu.edu.cn.

**Introduction**

Electrides represent a class of extraordinary compounds where some electrons in the solid are localized at interstitial regions, rather than being attached to atoms, and behave as anions[1,2]. The existence of quantized orbitals at the interstitials allows the transfer of electrons there[3]. Although the energies of both interstitial and atomic orbitals increase with pressure, the change of the interstitial orbital energy is usually smaller[4,5]. Thus, formation of electrides can become energetically favorable under high pressure, and various electrides have been already induced by pressure both in elements[6-9] and compounds[10-12]. These interstitial electrons largely determine the physical properties of the electrides[13-15].

Searching for high-temperature superconductivity remains one of the most important topics in condensed matter physics. Even though cuprates[16,17] and Fe-pnictides[18,19], among others, took a leading role on this pursuit, it is well known that pressure enhances the superconducting properties[20], and hydrogen sulfides[21] and lanthanum hydrides[22] have been recently observed to superconduct with $T_c$s above 200 K.

On the other hand, even though electrides are usually insulators, recent experiments showed that a canonical electride [Ca$_{24}$Al$_{28}$O$_{64}$]$^{4+}$(e$^-$)$_4$ becomes a superconductor ($T_c$ ~0.4 K)[23-26]. The interstitial electrons in Ca$_{24}$Al$_{28}$O$_{64}$ accommodate loosely in the cages of unit structures, rising an uncommon conductivity. Additionally, Mn$_5$Si$_3$-type Nb$_5$Ir$_3$[15] and two-dimension Y$_2$C[27] have been also found to hold both electride states and superconductivity. Additionally, as stated above, it is well known that pressure induces the formation of electrides. For example, both alkaline and alkaline-earth metals form electrides under pressure, as *s* orbital electrons can easily go to interstitial sites[4]. The strong localization of both interstitial and orbital electrons, caused by orbital coupling, makes them insulating[8,28-31]. However, under higher pressure Li shows a phase transition to a metallic phase while keeping the electride state (*Cmca*-24 Li at 90 GPa)[32]. Although this is a poor metal, the inclusion of extra elements might adjust interstitial electrons and help to improve its metallic character. For instance, as it was observed in suboxide $Li_6O$[33] and $Ca_2N$-type $Li_4N$[13] electrides, filling free spaces of Li with guest *p*-block elements modifies the electronic band topology and even increases its superconducting $T_c$[15]. Having this in mind, and considering that P has a moderate electronegativity and it is a remarkable superconductor[34,35], Li phosphides have reasonable expectances to become superconducting electrides. Interestingly, in a recent work a $Li_5P$ metallic electride has been found to be stable under pressure[36]. Li atoms in $Li_5P$ donate the excess electrons into the lattice spaces, forming an unusual 2D electride, with convenient conducting electronic channels.

These distinguishing electronic properties have attracted our attention of Li-P compounds. Herein, compounds with $Li_xP$ stoichiometry have been searched from 50 to 300 GPa, and several superconducting electrides (*i.e. C*2/*c* $Li_5P$, *P*-1 $Li_6P$, *C*2/*c* $Li_6P$ and *C*2/*c* $Li_8P$) have been found under pressure. In these electrides, electrons gather not only at interstitial sites but also around P atoms. Notably, the superconducting properties of the electrides are determined by their structural configurations ($Li_5P$, $Li_6P$ and $Li_8P$). *C*2/*c* $Li_6P$, with more connected interstitial electrons, shows the highest $T_c$ value of 39.3 K, which becomes the highest predicted $T_c$ between already reported electrides.

## Results and Discussion

### Ab initio structure prediction

Various Li-rich phosphides ($Li_xP$, $x$ = 1 - 8) have been searched extensively. Herein, structure prediction calculations were performed at 0 K and selected pressures of 50, 100, 200 and 300 GPa. The formation enthalpy ($\Delta H$), relative to the elemental solids (Li and P), was calculated at each considered pressure according to the equation below:

$$\Delta H(Li_xP) = [H(Li_xP) - xH(Li) - H(P)]/(x+1), \quad (1)$$

where $H = U + PV$ is the enthalpy of each composition, and $\Delta H$ is the formation enthalpy per atom of the given compound. Here, $U$, $P$, and $V$ are the internal energy, pressure, and volume, respectively. The relative thermodynamic stability of $Li_xP$ stoichiometries at 100, 200 and 300 GPa is shown in Figure 1a, and that at 50 GPa is in Figure S1. Stable phases lie on the global stability line of the convex hull, whereas compounds lying on the dotted lines are metastable with respect to decomposition into other $Li_xP$ compounds or elemental Li and P solids. In addition to the stable LiP (from 63 GPa up to 300 GPa), other Li-rich stoichiometries ($Li_3P$, $Li_5P$, $Li_6P$ and $Li_8P$) are predicted to be stable under increased pressure. The most stable *Fm*-3*m* $Li_3P$ lies on the convex hulls at the whole pressure range. Additionally, *P*6/*mmm* $Li_5P$ starts to be stable at 10.3 GPa, and remains stable until 163 GPa. There are two other $Li_5P$ stable phases with increasing pressure: *C*2/*c* $Li_5P$ is stable between 163 and 249 GPa, while *Cmcm* $Li_5P$ stabilizes above 249 GPa. For the stoichiometries with a higher Li content, $Li_6P$ and $Li_8P$ become stable over 178 GPa and 150 GPa, respectively. *P*-1 $Li_6P$ transfers to *C*2/*c* $Li_6P$ above 271 GPa. The pressure-composition phase diagram is shown in Figure 1b. All the phases in Figure 1b are dynamically stable, as they do not show any imaginary frequency modes (Figure S2).

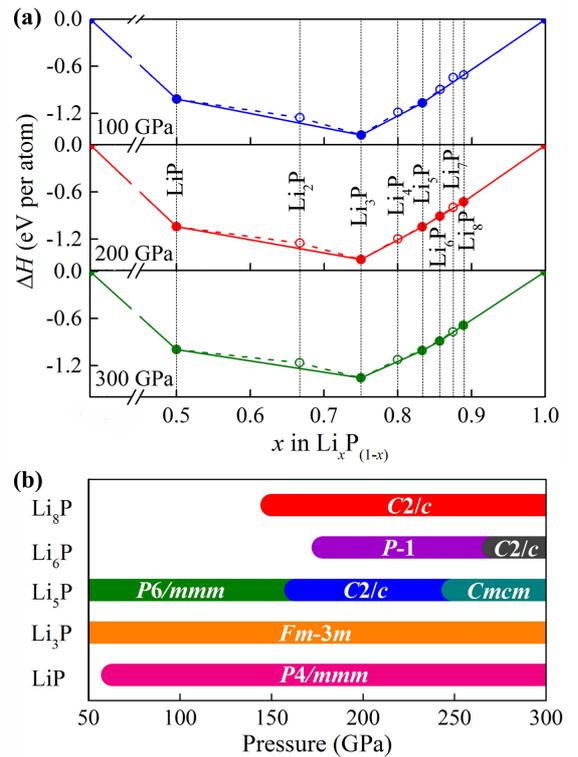

**Figure 1.** Relative thermodynamic stability of $Li_xP$ at 0 K and different pressures (100, 200 and 300 GPa). (a) The calculated formation enthalpy per atom of $Li_xP$ compounds with respect to elemental Li and P solids. The thermodynamically stable compounds are shown by solid symbols, connected by the solid lines line (convex hull). (b)

Pressure-composition phase diagram of Li$_x$P compounds.

**Geometry structures of stable Li$_x$P phases**

Pressure-induced LiP compound stabilizes into a tetragonal structure (space group *P*4/*mmm*, 1 formula unit, Figure 2a), in which each P atom has 8-fold coordination, with Li atoms forming a P-Li cuboid. As Li content increases, the coordination environment of the P atom changes and it becomes hypercoordinated. 15-fold and 17-fold coordinations of P atoms are observed in *C*2/*c* Li$_5$P and *Cmcm* Li$_5$P (Figure 2b and 2c), respectively; in *P*-1 Li$_6$P (Figure 2d) and *C*2/*c* Li$_8$P P atoms show a 16-fold coordination (Figure 2f). Notably, the highest coordination appears in *C*2/*c* Li$_6$P (4 formula units), in which P atoms coordinate with 18 nearest-neighbor Li atoms (Figure 2e). These coordination numbers are much higher than the previously calculated highest 14-fold coordination in *P*6/*mmm* Li$_5$P[36]. All Li-P phases are compactly assembled P-Li polyhedrons (a more detailed structural information can be found in the Supporting Information, Table S1).

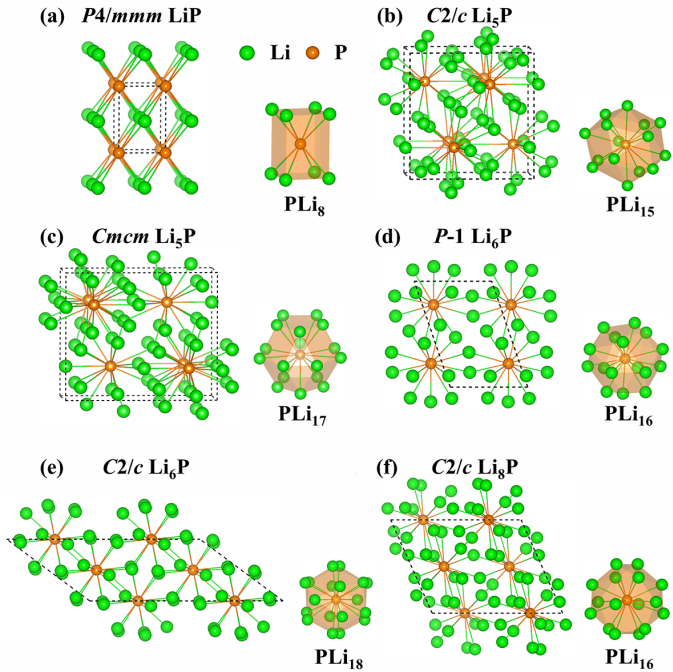

**Figure 2.** Structural features of stable Li-P phases at high pressures. (a) *P*4/*mmm* LiP at 200 GPa. (b) *C*2/*c* Li$_5$P at 200 GPa. (c) *Cmcm* Li$_5$P at 300 GPa. (d) *P*-1 Li$_6$P at 200 GPa. (e) *C*2/*c* Li$_6$P at 300 GPa. (f) *C*2/*c* Li$_8$P at 200 GPa. In all these structures, green and orange spheres represent Li and P atoms, respectively.

**Electronic properties of Li$_x$P**

Although P atom is hypercoordinated in all these Li-rich phosphide phases, Bader charge analysis does not conclude it has a hypervalence in any of them (Table S2). After getting three electrons from Li atoms, P atoms in Li-rich Li$_x$P ($x \geq 3$) fill completely their electronic shells. As a result, the rest of electrons offered by Li atoms accommodate into the lattice spaces, which induce the formation of electrides (Figures S3 and S6). The electronic band structures show that these electrides are metallic at the PBE level (*P*6/*mmm* Li$_5$P at 100 GPa, *C*2/*c* Li$_5$P at 200 GPa, *Cmcm* Li$_5$P at 300 GPa, *P*-1 Li$_6$P at 200 GPa, *C*2/*c* Li$_6$P at 300 GPa and *C*2/*c* Li$_8$P at 300GPa) (Figures S4 and 3c). Additionally, *P*4/*mmm* LiP, without interstitial electrons, is also metallic with a dominant P 3p orbital contribution (Figure S5). In contrast, *Fm*-3*m* Li$_3$P remains insulating until 300 GPa, as a result of a strong ionic bonding between Li and P, due to the high matching among 3Li$^+$ and P$^{3-}$ ions, which goes against closing the band.

As difference charge density in Figure 3a clearly shows, excess electrons in the highest coordinated *C*2/*c* Li$_6$P assemble into interstitials on the planes of phosphorus atoms, (0 1/3 0), having the dumbbell-like electride states (Figure S6). More interestingly, these dumbbell-like electride states interconnect each other across intermediate extranuclear electrons of P atoms, which is in sharp contrast with pressure induced isolated electride states in alkaline and alkaline-earth metals with insulating character[8]. This feature is further supported by its electron localization function (ELF)[37] (Figure. S6). These connected electronic channels in *C*2/*c* Li$_6$P favors the electronic conductivity, and the contribution of interstitial electrons to the metallic state is illustrated in the projected density of states (PDOS, Figure 3d). Although the main contribution at the Fermi level comes from Li 2p orbitals, it has an important contribution associated to the interstitial electrons. In contrast, as P atoms fill their electronic shells by attaining three electrons from Li atoms, P 3s and 3p orbitals barely reach the Fermi level. As shown in Figure S7, a well-defined Fermi surface nesting appears in Li$_6$P along Γ M (Figure 3b), with highly dispersive bands in this direction (Figure 3c). Conversely, flatter bands associated to more localized electronic states appear along Γ V and L A. As discussed above, nearly three electrons in *C*2/*c* Li$_6$P reside in the interstitials. In order to show this, we have built a hypothetical system by removing three electrons from $C2/c$ Li$_6$P, [Li$_6$P]$^{3+}$. The absence of interstitial electrons in both ELF and charge density difference of [Li$_6$P]$^{3+}$ (Figures S8a and S8b) confirm that these excess electrons are responsible for the electride states.

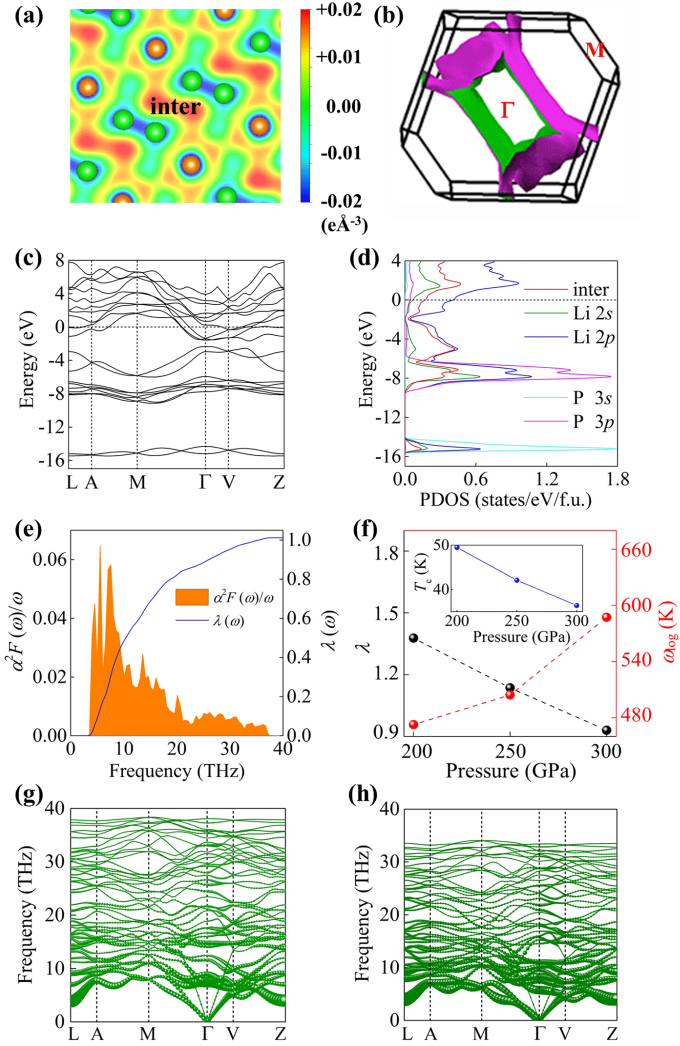

**Figure 3.** Electronic and superconducting properties of $C2/c$ Li$_6$P. (a) Difference charge density (crystal density minus superposition of isolated atomic densities) for $C2/c$ Li$_6$P plotted on the (0 1/3 0) plane at 300 GPa. (b) The nested Fermi surface along the Γ M direction for $C2/c$ Li$_6$P. The Fermi surface of each band crossing the Fermi energy is shown in Figure S7. (c) The electronic band structure of $C2/c$ Li$_6$P at 300 GPa. (d) PDOS of $C2/c$ Li$_6$P at 300 GPa. The red line labeled by "inter" is obtained by projecting onto the interstitial orbitals (interstitial-site-centered spherical harmonics in empty spheres with a Wigner-Seitz radius of 1.0 Å). (e) Eliashberg spectral function (orange area) and frequency-dependent electron-phonon coupling parameters $\lambda(\omega)$ (blue line) of $C2/c$ Li$_6$P at 270 GPa. (f) The electron-phonon coupling coefficient $\lambda$ (black dashed line) and the logarithmic average phonon frequency $\omega_{\log}$ (red dashed line) as a function of pressure. The critical temperatures $T_c$ (blue line) as a function of pressure is shown in the inset. The calculated phonon dispersion curves of $C2/c$ Li$_6$P at 300 (g), and 200 GPa (h). The area of each circle is proportional to the partial electron-phonon coupling, $\lambda_{q,v}$.

**Superconductivity in Li$_x$P**

Subsequently, we have calculated the superconducting $T_c$ of all the metallic Li-P phases as estimated from the McMillan-Allen-Dynes formula[38-40]:

$$T_C = \frac{\omega_{\log}}{1.2k_B}\exp\left[-\frac{1.04(1+\lambda)}{\lambda-\mu^*(1+0.62\lambda)}\right]. \quad (2)$$

Here, $k_B$ is the Boltzmann constant and $\mu^*$ is the Coulomb pseudopotential ($\mu^* = 0.1$). The electron-phonon coupling constant, $\lambda$, and the logarithmic average phonon frequency, $\omega_{\log}$, are calculated by the Eliashberg spectral function for electron-phonon interaction:

$$\alpha^2F(\omega) = \frac{1}{N(E_F)}\sum_{kq,v}|g_{k,k+q,v}|^2\delta(\varepsilon_k)\delta(\varepsilon_{k+q})\delta(\omega-\omega_{q,v}) \quad (3)$$

where $\lambda = 2\int d\omega \frac{\alpha^2F(\omega)}{\omega}$; $\omega_{\log} = \exp\left[\frac{2}{\lambda}\int\frac{d\omega}{\omega}\alpha^2F(\omega)\ln(\omega)\right]$.

Herein, $N(E_F)$ is the electronic density of states at the Fermi level, $\omega_{q,v}$ is the phonon frequency of mode $v$ and wave vector $q$, and $|g_{k,k+q,v}|$ is the electron-phonon matrix element between two electronic states with momenta $k$ and $k + q$ at the Fermi level[41,42].

**Table 1.** Superconducting properties of the metallic Li-P phases.

| Phases | Pressure (GPa) | $\lambda$ | $\omega_{\log}$ (K) | $N(E_f)$ (states/Ry) | $T_c$ (K) |
|---|---|---|---|---|---|
| $P4/mmm$ LiP | 200 | 0.28 | 898.11 | 2.40 | 0.22 |
| $C2/c$ Li$_5$P | 200 | 0.40 | 634.89 | 7.41 | 2.61 |
| $Cmcm$ Li$_5$P | 300 | 0.25 | 683.50 | 4.43 | 0.05 |
| $P$-1 Li$_6$P | 200 | 0.41 | 619.35 | 8.07 | 2.87 |
| $C2/c$ Li$_6$P | 270 | 1.01 | 554.34 | 11.49 | 39.30 |
| $C2/c$ Li$_8$P | 200 | 0.50 | 575.54 | 10.14 | 7.15 |

Main superconducting characteristics are shown in Table 1. Although metallicity extensively exists in high-pressure Li-P compounds, most of them present a low $T_c$, except $C2/c$ Li$_6$P, with a $T_c$ of 39.30 K at 270 GPa. Actually, it is the highest $T_c$ in an electride. The resulting electron-phonon coupling constant $\lambda$ of $C2/c$ Li$_6$P is 1.01 at 270 GPa, which is comparable to the maximum value for Li$_3$S (1.43)[43] and H$_2$P (1.04)[44]. As can be seen in Figure 3e, low-frequency phonon modes dominate superconductivity, so that modes below 10 THz give a contribution of 52.4% to the electron-phonon coupling parameter $\lambda$ at 270 GPa.

Although $C2/c$ Li$_6$P is metastable at pressures lower than 270 GPa, in order to better characterize the origin of its superconducting $T_c$, we have analyzed its evolution with pressure. As it is shown in Figure 3f, $T_c$ increases as pressure reduces (36.4 K at 300 GPa; 49.49 K at 200 GPa), with a pressure coefficient ($dT_c/dP$) of -0.13 K/GPa. In addition, while $\lambda$ decreases with pressure, $\omega_{log}$ increases. As it can be seen in Figure 3h, the contribution of low-frequency vibrations to the electron-phonon coupling is reduced with increasing of pressure. Interestingly, the softening along the Γ M direction decreases with pressure, which is correlated with the decreasing Fermi surface nesting. This nesting is associated with localized interstitial electrons, which are enhanced under lower pressure (Figure S9) and couple more strongly with phonons.

**Conclusion**

In this work we have analyzed pressure induced Li-rich phosphides under pressure up to 300 GPa. Although most of the electrides are non-metallic, in this work we report that pressure-induced stable Li$_5$P, Li$_6$P and Li$_8$P compounds are superconducting electrides. Among them, $C2/c$ Li$_6$P presents a $T_c$ of 39.3 K at 270 GPa, which is the highest among the already known electrides. Superconductivity in these compounds can be attributed to a combination of a weak electronegativity of P with a strong electropositivity of Li. Excess electrons in the highest coordinated $C2/c$ Li$_6$P assemble into the interstitials on the planes of phosphorus atoms forming dumbbell-like anionic electrons, which play a dominant role in the superconducting transition. A Fermi surface nesting associated to localized electronic bands induces a phonon softening that enhances the electron-phonon coupling and favors the superconducting transition. These results open up the interest to explore high-temperature superconductivity in similar binary compounds.

**Computational Method**

To search the thermodynamically stable candidates of Li-P alloys under pressure, we have employed the swarm-intelligence based CALYPSO structure prediction method, which can efficiently find the stable structures just depending on the given chemical compositions[45,46]. The CALYPSO method has made great achievements in predicting new compounds[22,47-50] Structural optimization and electronic property calculations were performed in the framework of density functional theory (DFT)[51,52] within the Perdew–Burke–Ernzerhof of the generalized gradient approximation (GGA)[53] as implemented in the VASP5.3 code[54]. The electron−ion interaction is described by pseudopotentials built within the scalar relativistic projector augmented wave (PAW)[55] method with $3s^23p^3$ valence electrons for P, and $1s^22s^12p^0$ valence electrons for Li. A cutoff energy of 500 eV and Monkhorst-Pack $k$-meshes[56] with a grid spacing of $2\pi \times 0.025$ Å$^{-1}$ were used to yield a good convergence for the enthalpy. To determine the dynamic stability of the predicted structures, phonon calculations were performed by using the finite displacement approach as implemented in the Phonopy code[57]. Superconducting properties were calculated based on density functional perturbation theory and the plane-wave pseudopotential method with Vanderbilt-type ultrasoft pseudopotentials, as implemented in the QUANTUM ESPRESSO code[58]. Detailed descriptions of structural predictions and computational details can be found in the Supporting Information.

**Acknowledgements**

This work is supported by the Natural Science Foundation of China under Nos. 21573037, 11704062, and 51732003; the Natural Science Foundation of Jilin Province (No. 20150101042JC); the Postdoctoral Science Foundation of China (under Grant No. 2013M541283); and the Fundamental Research Funds for the Central Universities (2412017QD006). A.B. acknowledges financial support from the Spanish Ministry of Economy and Competitiveness (FIS2016-76617-P) and the Department of Education, Universities and Research of the Basque Government and the University of the Basque Country (IT756-13).

# Supporting Information

# Superconductivity in Li$_6$P electride

Ziyuan Zhao[1], Shoutao Zhang[1], Tong Yu[1], Haiyang Xu[1], Aitor Bergara[*,2,3,4] and Guochun Yang[*,1]

[1]*Centre for Advanced Optoelectronic Functional Materials Research and Laboratory for UV Light-Emitting Materials and Technology of Ministry of Education, Northeast Normal University, Changchun 130024, China.*
[2]*Departmento de Física de la Materia Condensada, Universidad del País Vasco, UPV/EHU, 48080 Bilbao, Spain*
[3]*Donostia International Physics Center (DIPC), 20018 Donostia, Spain*
[4]*Centro de Física de Materiales (CFM), Centro Mixto CSIC-UPV/EHU, 20018 Donostia, Spain*

[*]Address correspondence to: a.bergara@ehu.eus; yanggc468@nenu.edu.cn.

# Computational Details

Our structural prediction approach is based on a global minimization of free energy surfaces merging *ab initio* total-energy calculations with CALYPSO (Crystal structure AnaLYsis by Particle Swarm Optimization) methodology as implemented in the CALYPSO code[1,2]. The structures of stoichiometry ($Li_xP$, $x$ = 1 - 8) were searched with simulation cell sizes of 1-4 formula units (f.u.) at 50, 100, 200 and 300 GPa. In the first step, random structures with certain symmetry are constructed in which atomic coordinates are generated by the crystallographic symmetry operations. Local optimizations using the VASP code[3] were done with the conjugate gradients method and stopped when Gibbs free energy changes became smaller than $1 \times 10^{-5}$ eV per cell. After processing the first generation structures, 60% of them with lower Gibbs free energies are selected to construct the next generation structures by PSO (Particle Swarm Optimization). 40% of the structures in the new generation are randomly generated. A structure fingerprinting technique of bond characterization matrix is applied to the generated structures, so that identical structures are strictly forbidden. These procedures significantly enhance the diversity of the structures, which is crucial for structural global search efficiency. In most cases, structural searching simulations for each calculation were stopped after generating 1000 ~ 1200 structures (e.g., about 20 ~ 30 generations).

To further analyze the structures with higher accuracy, we select a number of structures with lower enthalpies and perform structural optimization using density functional theory within the generalized gradient approximation[4] as implemented in the VASP code. The cut-off energy for the expansion of wavefunctions into plane waves is set to 500 eV in all calculations, and the Monkhorst–Pack *k*-mesh with a maximum spacing of $2\pi \times 0.025$ Å$^{-1}$ was individually adjusted in reciprocal space with respect to the size of each computational cell. This usually gives total energies well converged within ~1 meV/atom. The electron-ion interaction was described by using all-electron projector augmented-wave method (PAW) with $3s^23p^3$ valence electrons for P atom, $1s^22s^12p^0$ valence electrons for Li atom, respectively. To determine the dynamical stability of predicted structures, phonon calculations were performed by using the finite displacement approach as implemented in the Phonopy code.[5] In order to further test the reliability of the adopted pseudopotentials for Li and P, the validity of the projector augmented wave pseudopotentials from the VASP library is checked by comparing the calculated Birch-Murnaghan equation of state with that obtained from the full-potential linearized augmented plane-wave method (LAPW) using local orbitals (as implemented in WIEN2k[6]). The Birch-Murnaghan equation of states derived from PAW and LAPW methods are almost identical (Figure S0). Thus, our adopted pseudopotentials are feasible in the range of 50–300 GPa.

The electron-phonon coupling calculations are carried out with the density functional perturbation (linear response) theory as implemented in the QUANTUM ESPRESSO package.[7] We employ the ultrasoft pseudopotentials with $1s^22s^12p^0$ and $3s^23p^3$ as valence electrons for Li and P atoms, respectively. The kinetic energy cutoff for wave-function expansion is chosen as 40 Ry. To reliably calculate electron-phonon coupling in metallic systems, we need to sample dense $k$-meshes for electronic Brillouin zone integration and enough $q$-meshes for evaluating average contributions from the phonon modes. Dependent on specific structures of stable compounds, different $k$-meshes and $q$-meshes are used: 16 × 16 × 12 $k$-meshes and 4 × 4 × 3 $q$-meshes for LiP in the *P*4/*mmm* structure, 12 × 12 × 6 $k$-meshes and 4 × 4 × 3 $q$-meshes for Li$_5$P in the *C*2/*c* structure, 12 × 12 × 6 $k$-meshes and 4 × 4 × 3 $q$-meshes for Li$_5$P in the *Cmcm* structure, 12 × 12 × 8 $k$-meshes and 3 × 3 × 4 $q$-meshes for Li$_6$P in the *P*-1 structure, 12 × 12 × 6 $k$-meshes and 4 × 4 × 3 $q$-meshes for Li$_6$P in the *C*2/*c* structure, 15 × 15 × 8 $k$-meshes and 3 × 3 × 4 $q$-meshes for Li$_8$P in the *C*2/*c* structure.

# Supporting Figures

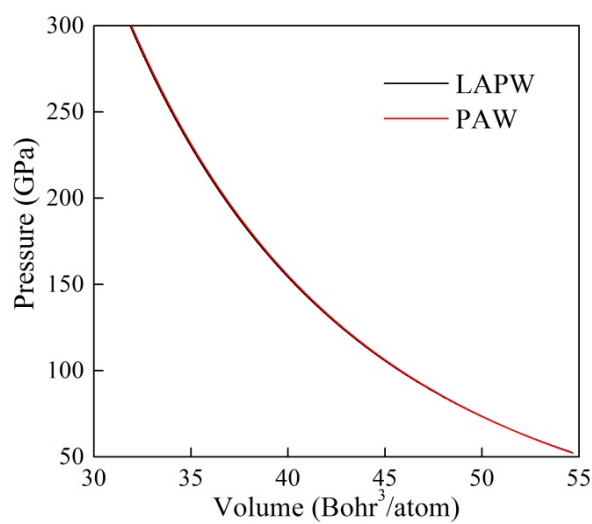

**Figure S0.** Comparison of the fitted Birch-Murnaghan equation of states for Li$_5$P with space group of *P*6/*mmm* by using the calculated results from the PAW pseudopotentials and the full-potential LAPW methods.

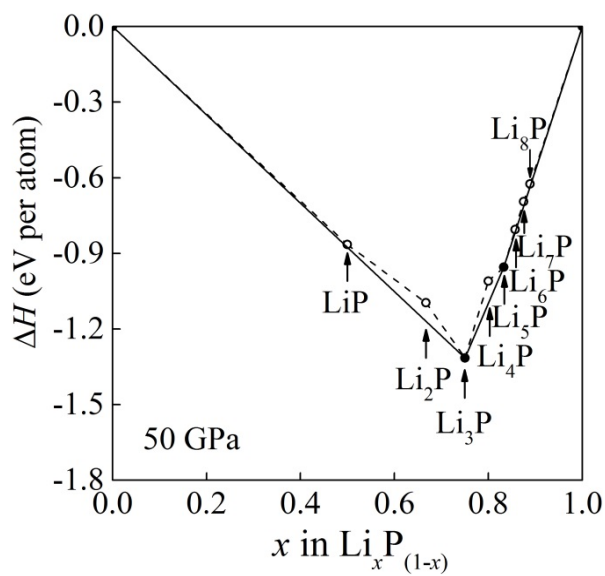

**Figure S1.** Relative thermodynamic stability of the Li$_x$P at 0 K and 50 GPa. The thermodynamically stable compounds are shown by solid symbols, connected by the convex hull line (solid lines).

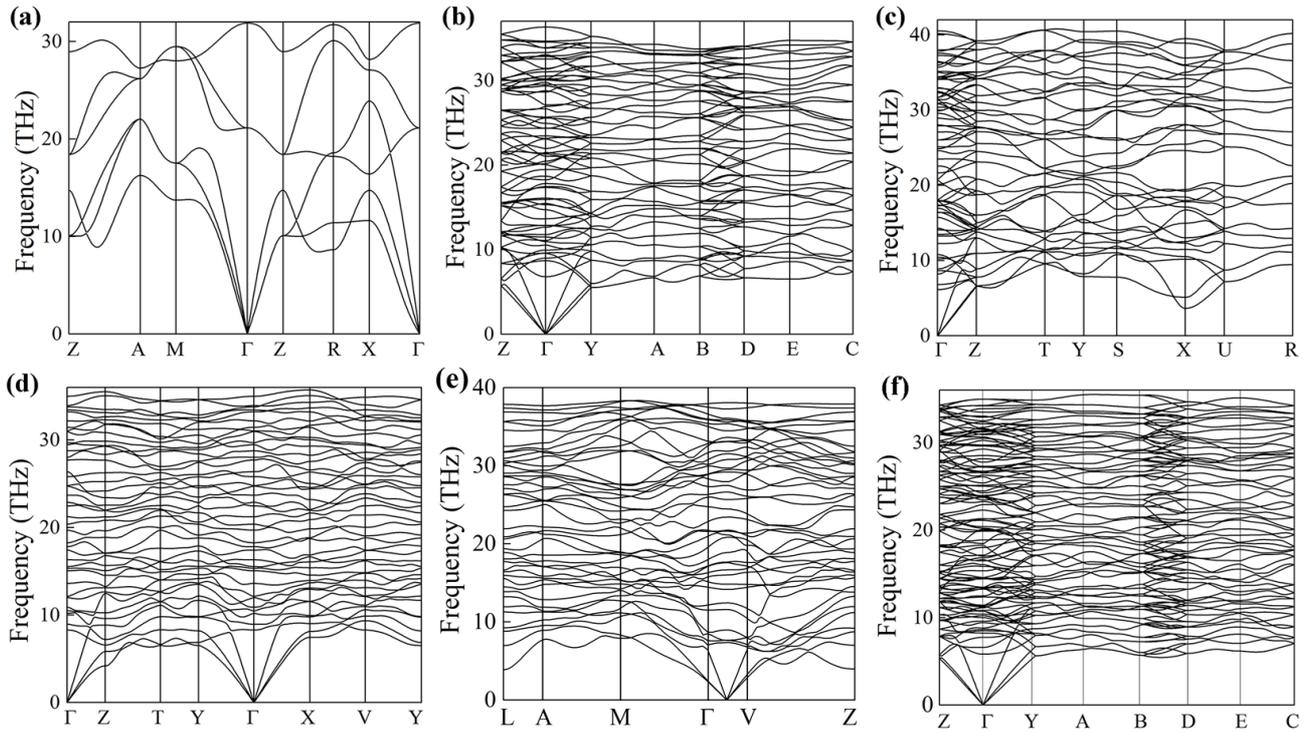

**Figure S2.** Phonon dispersion curves of metastable Li-rich phases. (a) *P*4/*mmm* LiP at 200 GPa. (b) *C*2/*c* $Li_5P$ at 200 GPa. (c) *Cmcm* $Li_5P$ at 300 GPa. (d) *P*-1 $Li_6P$ at 200 GPa. (e) *C*2/*c* $Li_6P$ at 300 GPa. (f) *C*2/*c* $Li_8P$ at 300 GPa. The absence of imaginary frequencies in these structures indicates they are dynamically stable.

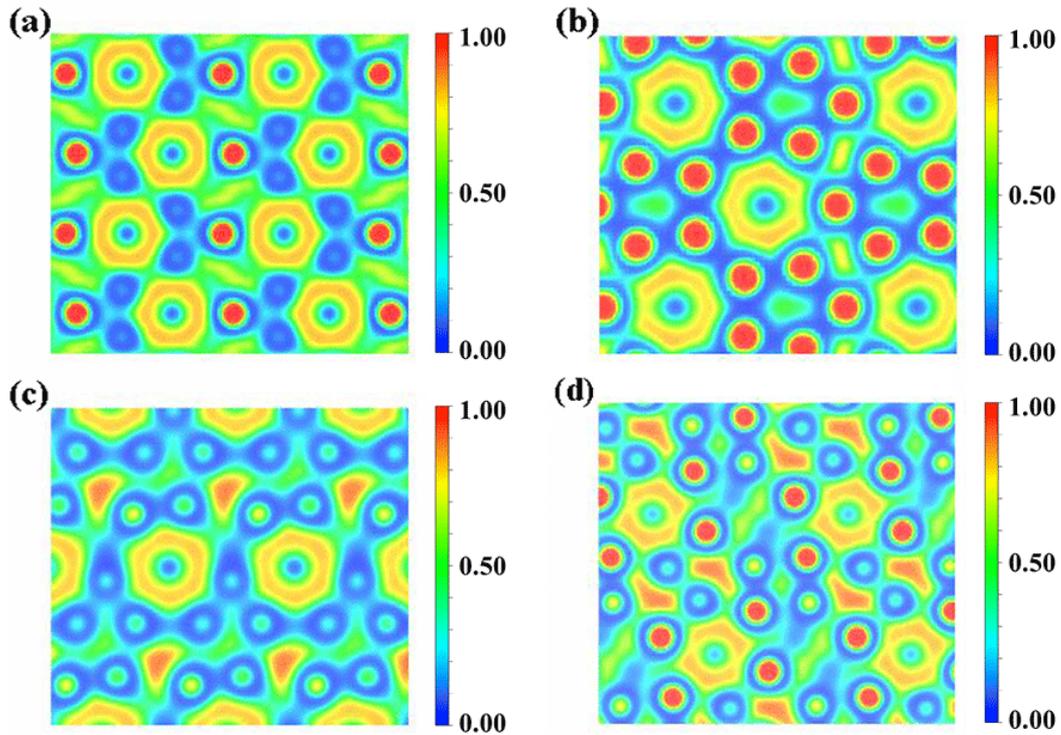

**Figure S3.** The calculated electron localization functions (ELF) for *C*2/*c* $Li_5P$ at 200 GPa (a), *Cmcm* $Li_5P$ at 300 GPa (b), *P*-1 $Li_6P$ at 200 GPa (c), and *C*2/*c* $Li_8P$ at 300 GPa (d).

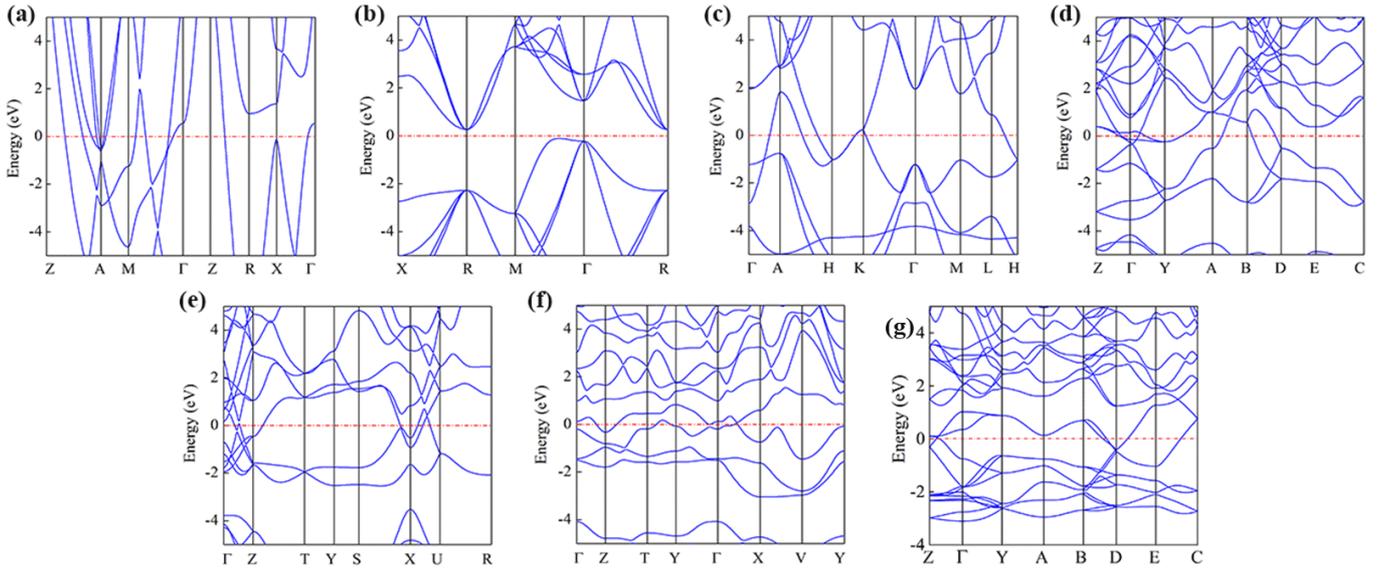

**Figure S4.** The electronic band structures have been calculated at the PBE level for *P4/mmm* LiP at 300 GPa (a), *Fm*-3*m* Li$_3$P at 300 GPa (b), *P6/mmm* Li$_5$P at 100 GPa (c), *C2/c* Li$_5$P at 200 GPa (d), *Cmcm* Li$_5$P at 300 GPa (e), *P*-1 Li$_6$P at 200 GPa (f), and *C2/c* Li$_8$P at 300GPa (g). All the high symmetric paths in Brillouin zone are selected from unit cells.

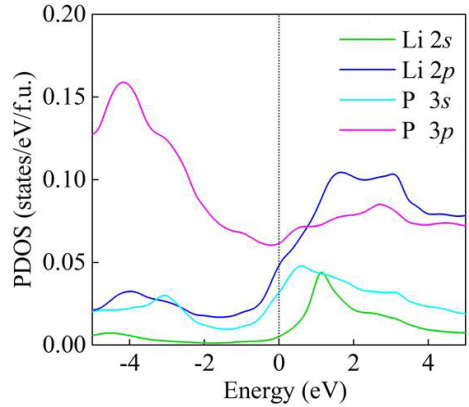

**Figure S5.** Projected density of states (PDOS) of *P4/mmm* LiP at 300 GPa.

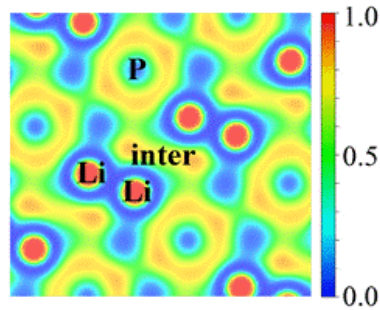

**Figure S6.** The calculated electron localization functions (ELF) for *C2/c* Li$_6$P on the (0 1/3 0) at 300 GPa. Electride states localize on the planes of phosphorus atoms, (0 1/3 0) and (0 2/3 0) planes, and their local features of the two mentioned planes are same. Therefore, only the calculated ELF on the (0 1/3 0) is exhibited.

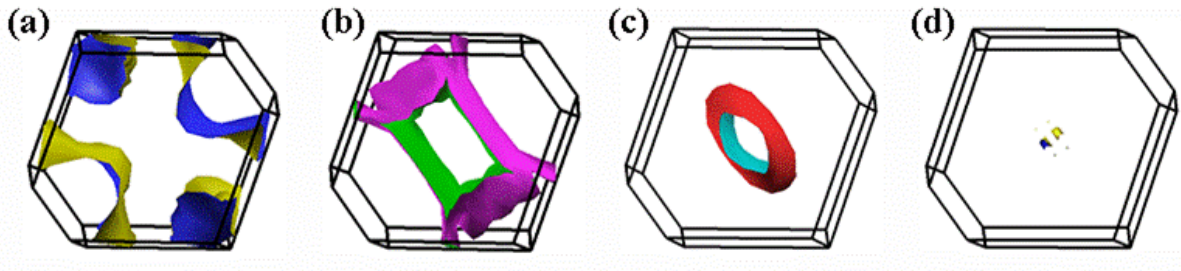

**Figure S7.** The Fermi surface associated to each band crossing the Fermi energy of *C*2/*c* Li$_6$P at 300 GPa.

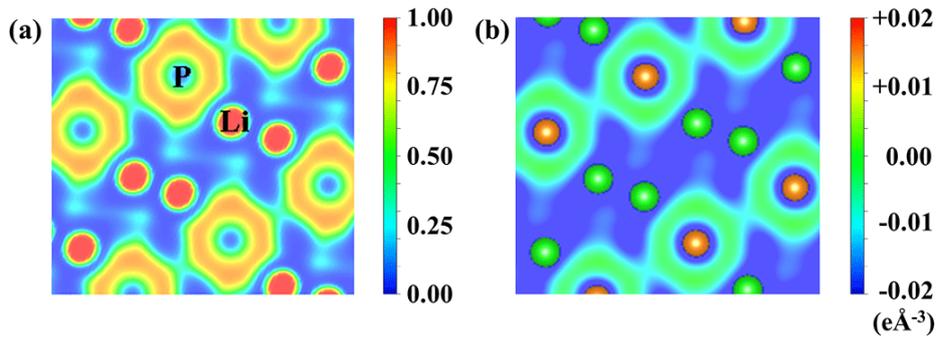

**Figure S8**. The ELF (a) and difference charge density (b) of [Li$_6$P]$^{3+}$ through removing 3 electrons per formula unit. The difference charge density of *C*2/*c* [Li$_6$P]$^{3+}$ with removing 3 electrons is obtained by subtracting the charge density of the isolated Li and the isolated P atom from that of *C*2/*c* Li$_6$P.

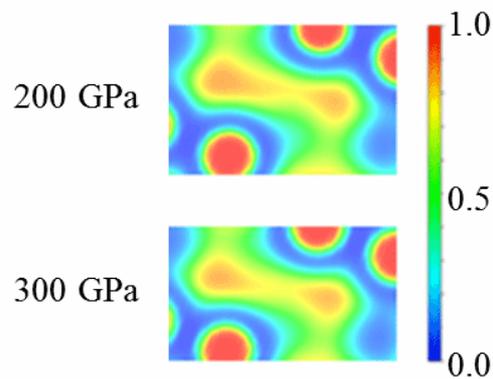

**Figure S9.** The ELF of interstitial electrons in *C*2/*c* Li$_6$P at 200 and 300 GPa.

# Supporting Tables

**Table S2.** Bader charge analysis of P atoms in Li-rich phases.

| Phases | The number of electrons obtained from Li per P atom |
|:---:|:---:|
| $C2/c$ Li$_5$P | 2.88 |
| $Cmcm$ Li$_5$P | 2.63 |
| $P$-1 Li$_6$P | 2.93 |
| $C2/c$ Li$_6$P | 3.00 |
| $Imma$ Li$_7$P | 2.65 |
| $C2/c$ Li$_8$P | 3.04 |